\documentstyle[12pt]{article}
\def\be{\begin{equation}}
\def\ee{\end{equation}}
\def\bea{\begin{eqnarray}}
\def\eea{\end{eqnarray}}

\begin{document}

\begin{center}
{\Large{\bf T-Duality and Mixed Branes }}                  
										 
\vskip .5cm   
{\large Davoud  Kamani}
\vskip .1cm
 {\it Institute for Studies in Theoretical Physics and 
Mathematics (IPM)
\\  P.O.Box: 19395-5531, Tehran, Iran}\\
{\sl e-mail: kamani@theory.ipm.ac.ir}
\\
\end{center}

\begin{abstract}

In this article the action of T-duality on a mixed brane is studied in the
boundary state formalism.
We also obtain a two dimensional mixed brane with non-zero electric
and magnetic fields, from a D$_1$-brane.

\end{abstract} 
\vskip .5cm

PACS:11.25.-w; 11.25.Mj; 11.30.pb 
\newpage
\section{Introduction}
 
It is well known that the 
T-duality in string theory is the target space duality, and generalizes 
the $R \rightarrow \alpha'/R$ duality in compactified string theory
\cite{1,2,3,4,5}. In fact, T-duality transformation is an exact symmetry
of closed string theory \cite{4}, and it can be used to relate problems
that might at first sight seem quite different.

In type II superstring theories, the action of T-duality
on $k$-directions (compactified on tori) 
of a D$_p$-brane produces a D$_{p-k}$-brane, and on 
$k$-directions, perpendicular to a D$_p$-brane, produces a D$_{p+k}$-brane.
Therefore for odd $``k"$, type IIA theory changes to the type IIB 
theory and vice-versa \cite{2,6,7}. In other words, the IIA theory 
compactified on a circle of radius $R$ is equivalent to the IIB theory
compactified on a circle of radius $\alpha'/R$ .

In this note we are interested in exploring 
what happens to a brane with internal 
back-ground field, under the action of the 
T-duality within the boundary state formalism. 
The boundary conditions 
corresponding to the closed string, emitted from
this brane, are combinations of Dirichlet and Neumann
boundary conditions, therefore we name it, ``mixed brane'', or
m$_p$-brane \cite{2,8,9,10}.

In section 2, we apply T-duality on the boundary state 
equations, corresponding
to a mixed brane. We shall see that the action of T-duality on two 
directions of certain mixed branes does not change their dimensions, unlike
D-branes. It instead changes the 
back-ground field strength. In section 3, with
the appropriate T-dualities we obtain a two dimensional mixed brane with all
field components, from a D$_1$-brane. The process can be generalized to
obtain higher dimensional branes with internal back-ground fields.
  
%%%%%%%%%%%%%%%%%%%%%%%%%%%%%%%%%%%%%%%%%%%%%%%%%%%%%%%%%%%%%%%%%%%%%%%%%%%%%
\section{T-duality on mixed boundary conditions} 
 
The boundary conditions for a closed string
emitted from a m$_p$-brane are \cite{2,8,9},

\bea
(\partial_\tau X^{\alpha}+ {\cal{F}}^{\alpha}_{\;\;\;\beta}
\partial_\sigma X^{\beta})_{\tau_0} \mid B \rangle \; = \;0 \;,
\eea
\bea
(\partial_{\sigma}X^i )_{\tau_0} \mid B \rangle \; = \; 0 \;,  
\eea
for the bosonic part, and
\bea
\bigg{(} (\psi^\alpha - i \eta {\tilde{\psi}}^{\alpha})
-{\cal{F}}^{\alpha}_{\;\;\;\beta} (\psi^\beta + i\eta 
{\tilde{\psi}}^{\beta})  \bigg{)}_{\tau_0}\mid B \rangle \; = \; 0 \;,  
\eea
\bea
(\psi^i + i \eta {\tilde{\psi}}^i)_{\tau_0}\mid B \rangle \; = \; 0 \;,
\eea
for the fermionic part, where $\tau_0$ is the $\tau$ variable on the
boundary of the closed string world sheet and $\eta=\pm1$. The
indices $\alpha$ and $\beta$ show the
brane directions and $i$ and $j$ show the directions perpendicular
to it. Total field strength is
\bea
{\cal{F}}_{\alpha \beta} = \partial_{[\alpha}A_{\beta]} 
-B_{\alpha \beta} \;,
\eea
where $A_\alpha$ is a $U(1)$ gauge field which lives on the brane, and 
$B_{\mu \nu}$ is the antisymmetric tensor of the NS$\otimes$NS sector of
the type II superstring
theories as back-ground, in the bulk of space-time.

Solution of the equations (1)-(4) gives the boundary state  
$\mid B \rangle$, which describes the mixed
brane \cite{8,9}.

Now consider T-duality in type II superstring theories. The closed string
mode expansion is
\bea
X^\mu (\sigma , \tau) = X^\mu _L (\tau+\sigma)   
+X^\mu _R (\tau-\sigma) \;,
\eea
\bea
X^\mu _L (\tau + \sigma)=x^\mu_L +2\alpha'p^\mu_L   
(\tau+\sigma) +\frac{i}{2} \sqrt{2\alpha'} \sum_{n \neq 0} \frac{1}{n}
{\tilde{\alpha}}^{\mu}_n e^{-2in(\tau+\sigma)} \;,
\eea
\bea
X^\mu _R ( \tau-\sigma )= x^\mu_R +2\alpha'p^\mu_R   
(\tau-\sigma) +\frac{i}{2} \sqrt{2\alpha'} \sum_{n \neq 0} \frac{1}{n}
{\alpha}^{\mu}_n e^{-2in(\tau-\sigma)} \;.
\eea
For non-compact direction $X^\mu$, we have $p^\mu_L = p^\mu_R =  
\frac{1}{2}p^\mu $, since $X^\mu$ is single valued at $\sigma$
and $\sigma+\pi$. If the direction $X^\mu$ is compact on a circle of radius
$R$, the left and the right components of the momentum are
\bea
p^\mu_L = \frac{1}{2\sqrt{\alpha'}} \bigg{(} \frac{\sqrt{\alpha'}}{R}M
+ \frac{R}{\sqrt{\alpha'}}N \bigg{)}\;,
\eea
\bea
p^\mu_R = \frac{1}{2\sqrt{\alpha'}} \bigg{(} \frac{\sqrt{\alpha'}}{R}M
- \frac{R}{\sqrt{\alpha'}}N \bigg{)}\;,
\eea
where the integers $N$ and $M$ are winding and momentum numbers of closed
string respectively.
Under the transformations
\bea
 \frac{R}{\sqrt{\alpha'}} \leftrightarrow  
\frac{\sqrt{\alpha'}}{R} \;\;\;\;,\;\;\;\;M \leftrightarrow N \;, 
\eea
the mass spectrum and interactions of the string theory are invariant.
The T-duality transformation on the compact direction $X^\mu$  can be written
\bea
T_{\mu}:\;\;X^\mu(\sigma , \tau) \rightarrow X'^\mu (\sigma , \tau)
=X^\mu_L - X^\mu_R \;,
\eea
which is a space-time parity transformation acting only on the right moving
degree of freedom. According to the (6)-(8), 
this gives transformations of (11) and 
\bea
\partial_{\tau}X^\mu \leftrightarrow \partial_\sigma X^\mu \;,
\eea
i.e. under the T-duality, the Dirichlet and Neumann boundary conditions get
exchanged. Worldsheet supersymmetry requires the following transformations
for the worldsheet fermions
\bea
T_\mu :\;\;\psi^\mu \rightarrow - \psi^\mu \;\;\;\;,\;\;\;\;
\tilde{\psi}^\mu \rightarrow  \tilde{\psi}^\mu \;;
\eea
these transformations cause the space-time 
metric $G_{\mu \nu}$ and field strength
${\cal{F}}_{\mu \nu}$ to transform under the T-duality, which can be
obtained by observing their action on the
NS$\otimes$NS sector oscillators
\bea
\mid G^{\mu \nu} \rangle = \bigg{(}b^\mu_{-1/2} {\tilde{b}}^{\nu}_{-1/2}+
b^\nu_{-1/2} {\tilde{b}}^{\mu}_{-1/2}\bigg{)}\mid 0 \rangle \;,
\eea
\bea
\mid B^{\mu \nu} \rangle = \bigg{(}b^\mu_{-1/2} {\tilde{b}}^{\nu}_{-1/2}-
b^\nu_{-1/2} {\tilde{b}}^{\mu}_{-1/2}\bigg{)} \mid 0 \rangle \;.
\eea
Under the transformations (14), we have $b^\mu_{-1/2} \rightarrow 
-b^\mu_{-1/2}$ and ${\tilde b}^\mu_{-1/2} \rightarrow 
{\tilde b}^\mu_{-1/2}$,
therefore,
\bea
T_\mu :
\left \{ \begin{array}{cl}
 G^{\mu \nu} \leftrightarrow 
-B^{\mu \nu}\;\;\;,\;\;\;\;\;\nu \neq \mu \\
\hspace{-1.8cm G^{\mu \mu} \rightarrow -G^{\mu \mu} \;.}\\
\end{array} \right.
\eea
We shall also observe from consistency that the transformations 
between $\eta_{\mu \nu}$ and ${\cal{F}}_{\mu \nu}$ have similar form. 

Now consider brane direction $X^{\alpha_c}$ which is assumed to 
be compact, therefore 
T$_{\alpha_c}$-duality acts on the boundary conditions (1)-(4) as
\bea
\left \{ \begin{array}{cl}
(\partial_\tau X^{\alpha'}+ {\cal{F}}^{\alpha'}_{\;\;\;\beta'}
\partial_\sigma X^{\beta'} )_{\tau_0} \mid B' \rangle \; = \;0 
\;\;\;\;,\;\;\;\; \alpha' , \beta' \neq \alpha_c \\
\hspace{-5.4cm (\partial_{\sigma}X^{\alpha_c} )_{\tau_0} 
\mid B' \rangle \; = \; 0 \;,} \\ 
\hspace{-5.5cm (\partial_{\sigma}X^i )_{\tau_0} 
\mid B' \rangle \; = \; 0 \;,}\\
\end{array} \right. 
\eea
\bea
\left \{ \begin{array}{cl}  
\bigg{(} (\psi^{\alpha'} - i \eta {\tilde{\psi}}^{\alpha'}) 
-{\cal{F}}^{\alpha'}_{\;\;\;\beta'} (\psi^{\beta'} + i\eta 
{\tilde{\psi}}^{\beta'})  \bigg{)}_{\tau_0}\mid B' \rangle \; = \; 0 \;,\\  
\hspace{-4cm (\psi^{\alpha_c} + i \eta {\tilde{\psi}}^{\alpha_c})_{\tau_0}
\mid B' \rangle \; = \; 0 \;,}\\ 
\hspace{-4.5cm (\psi^i + i \eta {\tilde{\psi}}^i)_{\tau_0}
\mid B' \rangle \; = \; 0 \;.}\\
\end{array} \right.
\eea
 We see that these are boundary state equations of 
 a mixed brane with field strength
${\cal{F}}_{\alpha' \beta'}$ and dimension $p-1$, i.e. 
$X^{\alpha_c}$ is perpendicular to this m$_{p-1}$-brane. The action
of T-duality on
one of the transverse compact directions of the m$_p$-brane, for example
$X^{i_c}$, gives a m$_{p+1}$-brane, i.e. $X^{i_c}$ is 
along this m$_{p+1}$-brane, which has field strength 
$\bar{\cal{F}}^{\bar \alpha}_{\;\;\; \bar \beta}$,
\bea
{\bar {\cal{F}}}^\alpha_{\;\;\;\beta}= {\cal{F}}^\alpha_{\;\;\;\beta}
\;\;\;\; , \;\;\;\; \bar{\cal{F}}^{\bar \alpha}_{\;\;\; i_c}=0 \;,
\eea
where $\{ {\bar \alpha} \}= \{i_c\} \bigcup \{ \alpha \}$.

If we simultaneously apply T-duality on two or more directions of a 
m$_p$-brane, some interesting effects appear. Consider two
compact directions of the brane, $X^{\alpha_0}$ and $X^{\beta_0}$.
From the states (15) and (16), we find that the metric and the 
antisymmetric tensor, under the T$_{\alpha_0 \beta_0}$-duality transform to
\bea
T_{\alpha_0 \beta_0}:
\left \{ \begin{array}{cl}  
\hspace{-3.5cm G^{\alpha_0 \beta_0} \rightarrow -G^{\alpha_0 \beta_0} \;,}\\ 
G^{\alpha_0 \nu'} \leftrightarrow -B^{\alpha_0 \nu'} \;\;\;\;,\;\;\;\;
\nu' \neq \alpha_0 , \beta_0 \;\;,\\
\hspace{-3.5cm B^{\alpha_0 \beta_0} \rightarrow 
-B^{\alpha_0 \beta_0} \;;} \\ 
\end{array} \right. 
\eea
therefore, the boundary state equations (1)-(4) transform to the following
forms,
\bea
\left \{ \begin{array}{cl} 
(\partial_\tau X^{\alpha'}+ {\cal{F}}^{\alpha'}_{\;\;\;\beta'}
\partial_\sigma X^{\beta'})_{\tau_0} 
\mid B'' \rangle \; = \;0 \;,\\
(\partial_\sigma X^{\alpha_0}- {\cal{F}}^{\alpha_0}_{\;\;\;\beta_0}
\partial_\tau X^{\beta_0} )_{\tau_0} \mid B'' \rangle \; = \;0 \;,\\
(\partial_\sigma X^{\beta_0}- {\cal{F}}^{\beta_0}_{\;\;\;\alpha_0}
\partial_\tau X^{\alpha_0} )_{\tau_0} \mid B'' \rangle \; = \;0 \;,\\
\hspace{-2.5cm (\partial_\sigma X^i)_{\tau_0} 
\mid B'' \rangle \; = \;0 \;,}\\
\end{array} \right. 
\eea
for the bosonic part, and
\bea
\left \{ \begin{array}{cl} 
\bigg{(} (\psi^{\alpha'} - i \eta {\tilde{\psi}}^{\alpha'})
-{\cal{F}}^{\alpha'}_{\;\;\;\beta'} (\psi^{\beta'} + i\eta 
{\tilde{\psi}}^{\beta'})  \bigg{)}_{\tau_0}\mid B'' \rangle \; = \; 0 \;,\\  
\bigg{(} (\psi^{\alpha_0} + i \eta {\tilde{\psi}}^{\alpha_0})
+{\cal{F}}^{\alpha_0}_{\;\;\;\beta_0} (\psi^{\beta_0} - i\eta 
{\tilde{\psi}}^{\beta_0})  \bigg{)}_{\tau_0}\mid B'' \rangle \; = \; 0 \;,\\  
\bigg{(} (\psi^{\beta_0} + i \eta {\tilde{\psi}}^{\beta_0})
+{\cal{F}}^{\beta_0}_{\;\;\;\alpha_0} (\psi^{\alpha_0} - i\eta 
{\tilde{\psi}}^{\alpha_0})  \bigg{)}_{\tau_0}
\mid B'' \rangle \; = \; 0 \;,\\  
\hspace{-4.6cm (\psi^i + i \eta {\tilde{\psi}}^i)_{\tau_0}
\mid B'' \rangle \; = \; 0 \;,}\\  
\end{array} \right. 
\eea
for the fermionic part 
,where $\alpha' , \beta' {\notin} \{ \alpha_0 , \beta_0\}$. From these
equations we conclude that,
if ${\cal{F}}^{\alpha_0}_{\;\;\;\beta_0}=0$, the
directions $X^{\alpha_0}$ and $X^{\beta_0}$ are perpendicular to the new
brane, (therefore it has dimension $p-2$). In the case  
${\cal{F}}^{\alpha_0}_{\;\;\;\beta_0} \neq 0 $, the 
dimension of the new brane is 
``$p$'' and field strength on it, is ${\cal{H}}^\alpha_{\;\;\;\;\beta}$,
where
\bea
\left \{ \begin{array}{cl} 
\hspace{-2.5cm {\cal{H}}^{\alpha'}_{\;\;\;\;\beta'}= 
{\cal{F}}^{\alpha'}_{\;\;\;\;\beta'} \;,}\\
{\cal{H}}^{\alpha_0}_{\;\;\;\;\beta_0}=
-{\cal{H}}^{\beta_0}_{\;\;\;\;\alpha_0}
=1/{\cal{F}}^{\alpha_0}_{\;\;\;\;\beta_0} \;,\\
\hspace{-1.6cm {\cal{H}}^{\alpha'}_{\;\;\;\;\alpha_0}=
{\cal{H}}^{\alpha'}_{\;\;\;\;\beta_0}=0 \;.}\\
\end{array} \right. 
\eea
The field strength ${\cal{H}}^\alpha_{\;\;\;\;\beta}$ is antisymmetric,
as expected. Similarly the action of the 
T-duality on three or more directions
of a mixed brane, changes it to another mixed brane.
%%%%%%%%%%%%%%%%%%%%%%%%%%%%%%%%%%%%%%%%%%%%%%%%%%%%%%%%%%%%%%%%%%%%%%%%%%%%%
\section{m$_2$-brane from D$_1$-brane}

Now  with appropriate actions of T-duality,
we obtain a m$_2$-brane with all the components of ${\cal{F}}$ that
are ${\cal{F}}_{01}$, ${\cal{F}}_{02}$ and ${\cal{F}}_{12}$, from a 
D$_1$-brane. Consider a  
D$_1$-brane parallel to the $X^1$-direction. The boundary state equations
of it, are
\bea
\left \{ \begin{array}{cl} 
(\partial_\tau X^0 )_{\tau_0} \mid B_d \rangle = 0 \;,\\ 
(\partial_\tau X^1 )_{\tau_0} \mid B_d \rangle = 0 \;,\\
(\partial_\sigma X^i )_{\tau_0} \mid B_d \rangle = 0 \;,\\
\end{array} \right. 
\eea
for the bosonic part, and
\bea
\left \{ \begin{array}{cl} 
(\psi^0 - i\eta {\tilde{\psi}}^0 )_{\tau_0} \mid B_d \rangle = 0 \;,\\ 
(\psi^1 - i\eta {\tilde{\psi}}^1 )_{\tau_0} \mid B_d \rangle = 0 \;,\\ 
(\psi^i + i\eta {\tilde{\psi}}^i )_{\tau_0} \mid B_d \rangle = 0 \;,\\ 
\end{array} \right. 
\eea
for the fermionic part, where $i \in \{2,3,...,9\}$. 

Consider ${\bar X}^1{\bar X}^2$-plane inside the $X^1X^2$-plane, which
${\bar X}^1$-direction makes angle $\theta$ with the $X^1$-direction, 
therefore
\bea
\left \{ \begin{array}{cl} 
X^1={\bar X}^1 \cos \theta - {\bar X}^2 \sin \theta \;,\\
X^2={\bar X}^1 \sin \theta + {\bar X}^2 \cos \theta \;,\\
\hspace{-2.9cm X^0={\bar X}^0 \;,}\\
\hspace{-2.9cm X^j={\bar X}^j \;,}
\end{array} \right. 
\eea
where $j \in \{3,4,...,9\}$.
From relations (27) and equations of motion of $\{X^\mu\}$,
we see that for ${\bar X}^\mu$ there
is $(\partial^2_{\tau}-\partial^2_{\sigma}){\bar X}^\mu = 0$,
which gives ${\bar X}^\mu = {\bar X}^\mu_L (\tau+\sigma)
+ {\bar X}^\mu_R (\tau-\sigma)$, therefore we can 
define  ${\bar{X'}}^\mu= {\bar X}^\mu_L- {\bar X}^\mu_R$. 
Let $X^1$ and $X^2$-directions be compact on circles of
radii R$_1$ and R$_2$. Therefore for appropriate
angle $\theta$ we can apply T-duality on ${\bar X}^1$ and 
${\bar X}^2$-directions. Since fermion components $\psi^\mu$ and  
${\tilde{\psi}}^\mu$ are space-time vectors, behaviors of them under the
space-time rotations are like $X^\mu$. Therefore, from now on we restrict 
ourselves to the bosonic part.

Application of relations (27) in equations (25) and then making 
T-duality on ${\bar X}^2$-direction give the
following boundary state equations (in the $\{ X^\mu\}$-coordinate system),
\bea
\left \{ \begin{array}{cl} 
\hspace{-3cm (\partial_{\tau} X^0 )_{\tau_0} \mid B' \rangle = 0 \;,}\\
(\cos \theta \partial_{\tau} X^1- \sin \theta \partial_{\sigma} X^2                  
)_{\tau_0} \mid B' \rangle = 0 \;,\\
(\cos \theta \partial_\tau X^2+ \sin \theta \partial_\sigma X^1                  
)_{\tau_0} \mid B' \rangle = 0 \;,\\
\hspace{-3cm (\partial_\sigma X^j)_{\tau_0} \mid B' \rangle = 0 \;;}\\
\end{array} \right. 
\eea
these equations describe a m$_2$-brane parallel to the $X^1X^2$-plane,
with the field strength $F_{12}=-F_{21}=-\tan \theta$ and
$F_{01} = F_{02} =0$ .

A m$_2$-brane with non-zero electric 
and magnetic fields can be obtained from 
the above m$_2$-brane by T-duality. A boost $v=\tanh \phi$, along
the direction $X^1$ gives the $\{ Y^\mu \}$-coordinate system,
\bea
\left \{ \begin{array}{cl} 
X^0=Y^0 \cosh \phi - Y^1 \sinh \phi \;,\\
X^1=-Y^0 \sinh \phi + Y^1 \cosh \phi \;,\\
\hspace{-3.5cm X^2=Y^2 \;,}\\
\hspace{-1cm X^j=Y^j\;\;\;,\;\;\;j \neq 0,1,2 \;.}
\end{array} \right. 
\eea
In this system, the components of field strength are 
${\cal{H}}^\alpha_{\;\;\;\beta}$ , where
\bea
{\cal{H}}^0_{\;\;\;1} = 0\;\;\;,\;\;\; {\cal{H}}^0_{\;\;\;2} =
-\sinh \phi \tan \theta \;\;\;,\;\;\;
{\cal{H}}^1_{\;\;\;2} =- \cosh \phi \tan \theta \;,
\eea
therefore
\bea
F^1_{\;\;\;2} = -\sinh \phi {\cal{H}}^0_{\;\;\;2} 
+ \cosh \phi {\cal{H}}^1_{\;\;\;2} \;.
\eea
Assume that the time direction $X^0$ also be compact on circle of radius
$R_0$. For appropriate values of $\phi$, 
the coordinates $Y^0$ and $Y^1$ also are periodic.

Introducing relations (29) and (31) in equations (28), and then making 
T-duality, simultaneously on 
$Y^0$ and $Y^2$ directions (use the transformations (21) to transform
field strength ${\cal{H}}$), give the
boundary state equations of the form (1) and (2) (in the
$\{ X^\mu \}$-coordinate system) with the field strength,
\bea
\left \{ \begin{array}{cl} 
{\cal{F}}^0_{\;\;\;1} = {\cal{F}}^1_{\;\;\;0} = -\coth \phi \;,\\ 
{\cal{F}}^0_{\;\;\;2} = {\cal{F}}^2_{\;\;\;0} = -\frac{\coth \phi} 
{\tan \theta \sinh^2 \phi} \;,\\
{\cal{F}}^1_{\;\;\;2} =-{\cal{F}}^2_{\;\;\;1} =  
\frac{1}{\tan \theta \sinh^2 \phi} \;.\\
\end{array} \right. 
\eea
Note that according to the space-time metric $\eta_{\mu \nu} =
{\rm diag}(-1,1,...,1)$, the field strength 
${\cal{F}}^\alpha_{\;\;\;\beta}$ is
antisymmetric. 

We obtained a m$_2$-brane which contains non-zero electric
and magnetic fields from a D$_1$-brane. More applications of T-duality,
on the appropriate directions, lead to the mixed branes with 
higher dimensions.

Similarly application of T-duality on a direction makes angle with a 
D$_p$-brane, produces a m$_{p+1}$-brane with magnetic field, and on some 
directions (including the time coordinate) of the boosted system, produces
electric field on mixed brane.
%%%%%%%%%%%%%%%%%%%%%%%%%%%%%%%%%%%%%%%%%%%%%%%%%%%%%%%%%%%%%%%%%%%%%%%%%%%%
\section{Conclusion}
					 
By using the boundary state formalism, we saw that
the action of T-duality on one direction of a m$_p$-brane, changes it to a 
m$_{p-1}$-brane, and on $k$-directions perpendicular to the m$_p$-brane,
changes it to a m$_{p+k}$-brane, and on some pair of directions of a
certain mixed brane, does not change the dimension of it, this action
changes the mixed brane to another mixed brane with the modified field
strength. 

Application of the T-duality on appropriate directions, 
changes a D$_p$-brane to a mixed brane. By successive applications of 
the T-duality, we specially obtained a m$_2$-brane
from a D$_1$-brane.

{\bf Acknowledgement}

The author would like to thank H. Arfaei for useful discussion.

%%%%%%%%%%%%%%%%%%%%%%%%%%%%%%%%%%%%%%%%%%%%%%%%%%%%%%%%%%%%%%%%%%%%%%%%%%%%

\end{document}